\documentclass[
preprint,
superscriptaddress,
amsmath,amssymb,
aip,
showpacs,
]{revtex4-1}

\usepackage{graphicx}
\usepackage{dcolumn}
\usepackage{bm}

\newcommand {\SNO}{Sn$_2$Nb$_2$O$_7$}  
\newcommand {\STO}{Sn$_2$Ta$_2$O$_7$}  
\newcommand {\PNO}{Pb$_2$Nb$_2$O$_7$}  
\newcommand {\PTO}{Pb$_2$Ta$_2$O$_7$}  
  
\newcommand {\ABO}{$A_2B_2$O$_7$}  

\newcommand {\FBO}{$A_2B_2$O$_7$ (\textit{A} = Sn, Pb; \textit{B} = Nb, Ta)}  

\usepackage{hyperref}
\hypersetup{
setpagesize=false,
 bookmarksnumbered=true,%
 bookmarksopen=true,%
 colorlinks=true,%
 linkcolor=blue,
 citecolor=blue,
 urlcolor=blue
}

\setcitestyle{numbers,square}  

\makeatletter
\newcommand{\colorcaption}[2][]{%
	\begingroup%
	\renewcommand{\@caption@fignum@sep}{ (Color online). }%
	\caption[#1]{#2}%
	\endgroup%
}
\makeatother
\begin{document}

\preprint{APS/123-QED}

\title{Trends in Bandgap of Epitaxial \textit{A}$_2$\textit{B}$_2$O$_7$ (\textit{A} = Sn, Pb; \textit{B} = Nb, Ta) Films Fabricated by Pulsed Laser Deposition}

\author{T. C. Fujita}%
\email{fujita@ap.t.u-tokyo.ac.jp}
\author{H. Ito}
 \affiliation{
 Department of Applied Physics and Quantum Phase Electronics Center, University of Tokyo, Tokyo 113-8656, Japan
}%

\author{M. Kawasaki}
\affiliation{ 
	Department of Applied Physics and Quantum Phase Electronics Center, University of Tokyo, Tokyo 113-8656, Japan
}%
\affiliation{ 
	RIKEN Center for Emergent Matter Science (CEMS), Wako 351-0198, Japan
}

\date{\today}

\begin{abstract}
Pyrochlore oxides {\ABO}  have been a fruitful playground for condensed matter physics because of the unique geometry in the crystal structure.
Especially focusing on the \textit{A}-site tetrahedral sub-lattice in particular pyrochlore oxides {\FBO}, recent theoretical studies predict the emergence of ``quasi-flat band'' structure as a result of the strong hybridization between filled \textit{A}-n\textit{s} and  O-2\textit{p} orbitals. 
In this work, we have established the growth conditions of {\SNO}, {\STO}, {\PNO}, and {\PTO} films by pulsed laser deposition on Y-stabilized ZrO$_2$~(111) substrates to elucidate their optical properties. 
Absorption-edge energies, both for direct and indirect band gaps, increase in the order of {\SNO}, {\STO}, {\PNO}, and {\PTO}.
This tendency can be well explained by considering the energy level of the constituent elements. 
Comparison of the difference between direct and indirect band gaps reveals that Pb$_2$$B_2$O$_7$ tends to have a less dispersive valence band than Sn$_2$$B_2$O$_7$.
Our findings are consistent with the theoretical predictions and are suggestive of the common existence of the hybridized states in this class of compounds.
\end{abstract}
\maketitle

\begin{figure}[h]
	\includegraphics[width=5.7cm]{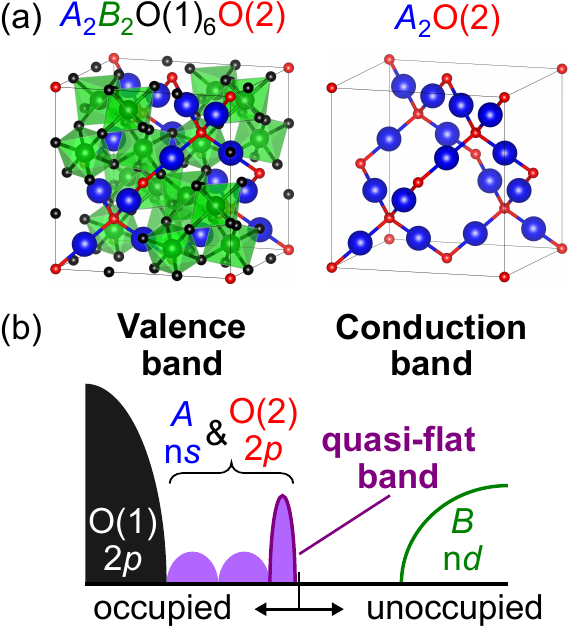}
	\colorcaption{\label{concept}
		(a) Crystal structure of a pyrochlore oxide ({\ABO} = $A_2B_2$O(1)$_6$O(2)) and its tetrahedral $A_2$O(2) sub-lattice that forms upper part of the valence band.  Schematic illustrations are drawn by using VESTA~\cite{VESTA}.
		(b) Schematic band diagram of pyrochlore oxides with ``quasi-flat band'' structure. 
	}
\end{figure}

Pyrochlore oxides are one of the typical complex oxides generally formulated as {\ABO} and are known to exhibit a wide range of physical properties~\cite{SUBRAMANIAN1983} such as ferro-electricity~\cite{Jona1955},  superconductivity~\cite{Hanawa2001,Sakai2001}, and frustrated magnetism~\cite{Ramirez1999,Gardner2010}.
This crystal structure can be also denoted by $A_2B_2$O(1)$_6$O(2) when distinctive two oxygen sites are explicitly formulated.
O(1)  sites form $B$O(1)$_6$ octahedra and O(2) sites configurate $A_2$O(2) tetrahedral network as shown in Fig.~\ref{concept}(a).  
Recent theoretical studies predict that this $A_2$O(2) tetrahedral network plays a crucial role in the emergence of ``quasi-flat band" structure in particular pyrochlore oxides {\FBO} ~\cite{Hase2018_1, Hase2018_2, Hase2019,Hase2020}.
Different from ordinary {\ABO}, where \textit{A}-site element does not contribute to the band structure near Fermi level, these compounds have hybridization of filled Sn-5\textit{s} (or Pb-6\textit{s})  and O-2\textit{p} orbitals forming the valence band maximum (VBM) as shown in Fig.~\ref{concept}(b). 
This VBM is not completely flat but flat enough in the sense that the bandwidth is much narrower than the energy scale of the band gap, which is the reason why it is called quasi-flat band. 
On the other hand, conduction band minimum (CBM) consists of empty Nb-4\textit{d} (or Ta-5\textit{d}) orbital and is far from the Fermi level. 
Because of this unique band structure, this class of compounds is expected to be an ideal platform for realizing numbers of emergent physical properties related with flat band structure, such as ferro-magnetism, high-temperature superconductivity,  the fractional quantum Hall effect~\cite{Liu2014}, and topological states~\cite{Zhang2019,Zhou2019,Hase2020}.


According to the theoretical band calculations, band structure near Fermi level largely depends on the combination of \textit{A}- and \textit{B}-site elements~\cite{Hase2018_1,Hase2018_2,Hase2019,Hase2020}.
We have already reported an optical band gap of {\SNO} epitaxial thin films and its modulation by  Ti-doping in an attempt to dope holes in the quasi flat band~\cite{Ito2021}.  
It has been revealed that the hybridized Sn-5\textit{s} and O-2\textit{p} orbitals do contribute to the band structure near Fermi level, which is consistent with the theoretical predictions. 
In this aspect, it is gripping to examine the optical properties of other candidate materials that possibly host the quasi-flat band structures. 
Also, from the materials point of view, it is important to establish the growth conditions for oxides containing highly volatile elements such as tin and lead.  
Here, in addition to {\SNO}, we report epitaxial thin film growth of  {\STO}, {\PNO}, and {\PTO}, and their optical properties.

Epitaxial  {\FBO} thin films  were prepared on Y-stabilized ZrO$_2$ (YSZ) (111) substrates by a pulsed laser deposition.
Before the film growth, YSZ substrates were annealed in air with an electronic furnace at 1,350$ \ {}^\circ\mathrm{C}$ for 3 hours to obtain a clear step-terrace structure with single-unit-cell height ($\sim$3 \AA).
The targets were prepared by a hot press method in 400 Torr Ar atmosphere under a mechanical pressure of 80 MPa, starting from a mixture of SnO$_2$, PbO, Nb$_2$O$_5$, and Ta$_2$O$_5$ powders.
Chemical composition of the mixed powder for Sn$_2$$B_2$O$_7$ targets was Sn:\textit{B} = 1:1, while that for Pb$_2$$B_2$O$_7$ was Pb:\textit{B} = 1.1:1.
The sintering temperature was 1,100$ \ {}^\circ\mathrm{C}$ for Sn$_2$$B_2$O$_7$  and 850$ \ {}^\circ\mathrm{C}$ for Pb$_2$$B_2$O$_7$.
The films were deposited at various growth temperatures and O$_2$ pressures by KrF excimer laser ($\lambda$= 248 nm) pulses at a frequency of 5 Hz. 
Structural properties of the films were characterized by x-ray diffraction (XRD) (Smart Lab, Rigaku) at room temperature. 
Chemical compositions of the films were examined by energy dispersive X-ray spectroscopy with scanning electron microscope (SEM-EDX), for which the electron beam was aligned normal to the film surface. 
Optical properties were measured by UV-3600 (SHIMADZU) at room temperature.

\begin{figure}
	\includegraphics[width=6.8cm]{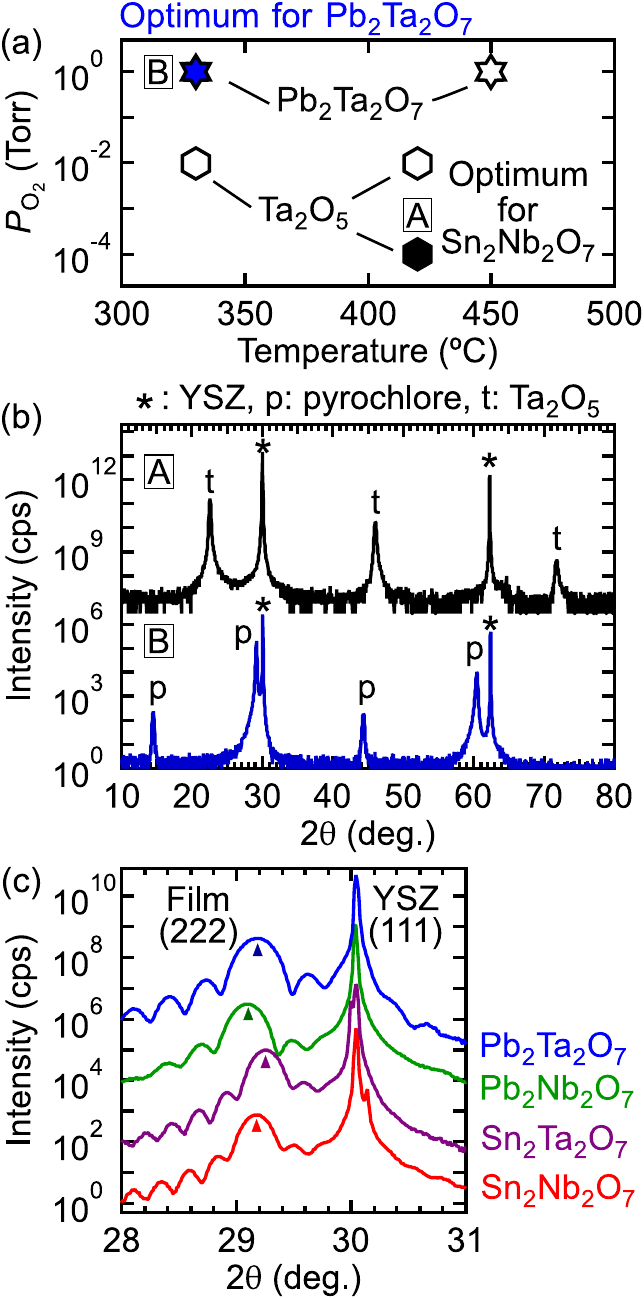}
	\colorcaption{\label{XRD} 
		(a) Growth phase diagram of {\PTO} as functions of oxygen partial pressure ($P_{\mathrm{O}_2}$) and growth temperature.  The phases appearing in the films are indicated by hexagons and stars for hexagonal Ta$_2$O$_5$  and pyrochlore {\PTO}, respectively. 
		A is the reported optimized conditions for {\SNO} (Ref.~\cite{Ito2021}) and can be applied for {\STO} as well. 
		B is the optimized conditions for {\PTO} and also applied for {\PNO} in this work.
		Note that target-substrate distance and laser power are also optimized (See main text).
		(b) 2$\theta-\omega$ XRD scans for the films grown under the conditions denoted as A and B in (a). Peaks of YSZ substrate,  {\PTO}, and hexagonal Ta$_2$O$_5$ are denoted by asterisk, p, and t, respectively. 
		(c) 2$\theta-\omega$ XRD scans around (222) peak of {\SNO},  {\STO}, {\PNO}, and {\PTO}  films prepared under optimized conditions.
		Thickness of the films are 40, 40, 35, and 32 nm, respectively. 
		Peak positions of the films are denoted by filled triangles. 		
	}
\end{figure}

We first discuss the optimization of the thin film growth conditions. 
Following the growth conditions for {\SNO} reported previously~\cite{Ito2021}, we are able to obtain high-quality {\STO} films at the growth temperature of 420$ \ {}^\circ\mathrm{C}$ under 10$^{-4}$ Torr O$_2$.
These conditions are denoted as A in Fig.~\ref{XRD}(a). 
However, in the case of {\PTO} films, only hexagonal Ta$_2$O$_5$ phase appears under the same conditions (top panel of Fig.~\ref{XRD}(b)).  
The same (absence of \textit{A}-site element and appearance of hexagonal $B_2$O$_5$ binary oxide) has occurred for Sn$_2$Nb$_2$O$_7$ as well, when the growth temperature is higher as a result of the high volatility of SnO~\cite{Ito2021}.
Considering that the vapor pressure of PbO is higher than that of SnO~\cite{Lamoreaux1987}, we can expect that lower growth temperature and higher O$_2$ pressure are required to suppress the re-evaporation of PbO. 
Indeed, pure Pb$_2$Ta$_2$O$_7$ phase appears at growth temperatures of 330 and 450$ \ {}^\circ\mathrm{C}$ under 1 Torr O$_2$ (Fig.~\ref{XRD}(a)).
Especially at 330$ \ {}^\circ\mathrm{C}$, higher quality Pb$_2$Ta$_2$O$_7$ film is obtained as shown in the bottom panel of Fig.~\ref{XRD}(b).
These conditions are denoted as B in Fig.~\ref{XRD}(a), and high-quality {\PNO} film is also obtained at this point.
Here, it is worth noting that in order to compensate the short mean-free-path of the precursors under high O$_2$ pressure ($\sim7$ mm at 1 Torr), the distance between the target and substrate is set as narrow as 15 mm in the fabrication of  Pb$_2$$B_2$O$_7$ and laser fluence is 2.2 J/cm$^2$, while they are 25 mm and 1 J/cm$^2$, respectively, for Sn$_2$$B_2$O$_7$.

Figure~\ref{XRD}(c) shows XRD $2\theta$-$\omega$ scan around (222) peak of {\SNO},  {\STO}, {\PNO}, and {\PTO}  films prepared under the optimized conditions discussed above.
Laue oscillations are clearly observed around the (222) main peak for all the films, indicating sharp interfaces and smooth surfaces of the films. 
Atomic micro scope images also show the smooth surface of the films with route-mean-square roughness of $\sim$1 nm or smaller (Fig. S2).
The full width at half maximum (FWHM) of the rocking curves around (222) peaks is less than 0.07$^\circ$ for all the films (not shown), evidencing the high quality of the films as well.
Reciprocal space mappings (RSM) around pyrochlore (662) peaks reveal that all the films are almost fully relaxed (Fig.~S1(a)), which is expected by the rather large lattice mismatch (2--5 \%) between YSZ  substrate and the films as summarized in Table S1.
Azimuthal ($\phi$) scans around  pyrochlore (662) show three-fold symmetry, which is the same as that of the YSZ substrate (331), indicating that in-plane misoriented domains are indiscernible in these films (Fig.~S1(b)).
There has been reported a paper on reflection high energy electron diffraction (RHEED) intensity oscillation while growing thin films of a pyrochlore Tb$_2$Ti$_2$O$_7$ at rather high growth temperature (750$ \ {}^\circ\mathrm{C}$)~\cite{bovo2017}.
In our case, we did not observe RHEED oscillation, presumably due to rather low growth temperatures set with considering highly volatile elements of Sn and Pb.

The chemical composition of the films is estimated by SEM-EDX.
Because Nb peak is overlapped by Zr strong peak originating from YSZ substrate, the exact composition of cations in {\SNO} and  {\PNO} films cannot be measured.
However,both {\STO} and {\PTO} films exhibit the stoichiometric cation ratio  (Sn (or	Pb)/Ta $\approx$ 1) within the accuracy of our measurements ($\pm5\%$) in this study.
Therefore, we can fairly postulate that cation off-stoichiometry is negligible in our films.
Besides the cation ratio, anti-site defects Sn$^{2+}$/Sn$^{4+}$ (or Pb$^{2+}$/Pb$^{4+}$), and oxygen vacancies are also important in terms of the electronic structures of pyrochlore oxides and are indeed discussed in several previous works with bulk samples especially for {\SNO}~\cite{Aiura2017,Kikuchi2017,Samizo2019, Samizo2021}. 
However, quantitative evaluation of those is quite challenging in thin-film samples, and thus would be left for future research. 

\begin{figure}[h]
	\includegraphics[width=6.9cm]{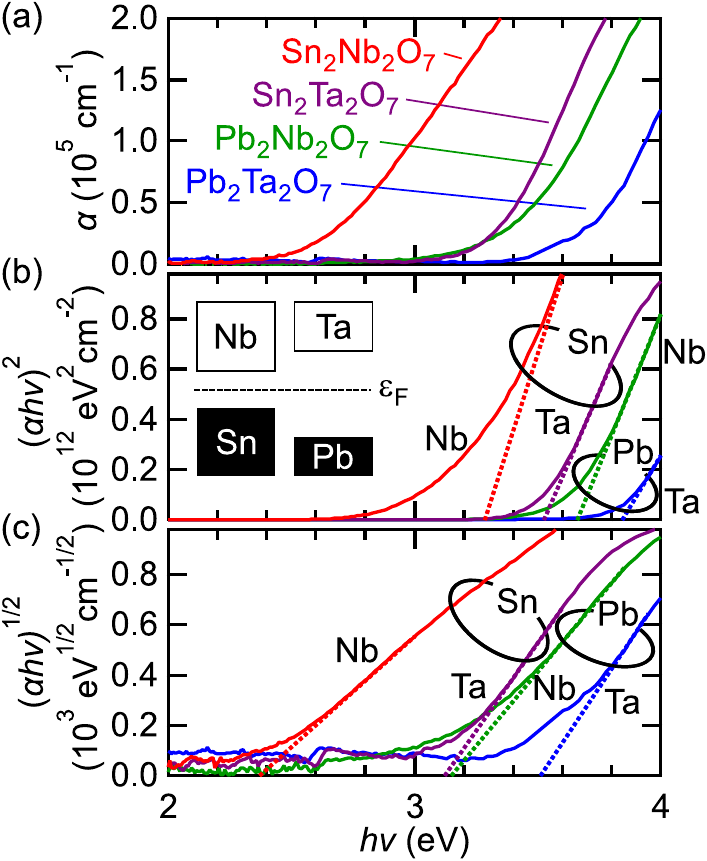}
	\colorcaption{\label{opt} 
		(a) Absorption constant ($\alpha$) for the same set of {\SNO},  {\STO}, {\PNO}, and {\PTO}  films as these in Fig.~\ref{XRD}(c).  
		Tauc plot of $\alpha$ assuming the direct (b) and indirect (c) transitions.
		Schematic energy levels of each constituent cation are shown in the inset of (b).
			}
\end{figure}

We now argue the optical properties of the films to characterize the band structures in this class of compounds.
Absorption coefficient ($\alpha$) values were calculated as below: 
\begin{equation}
\alpha = -\frac{1}{d} \mathrm{ln}\frac{T}{1-R}
	\label{eq:alpha}
\end{equation}
where $d$ is the film thickness, $T$ is the fraction of incident light  transmitted through the film, and $R$ is the fraction of incident light reflected by the film.
The absorption spectra of the films are presented in Fig.~\ref{opt}(a). 
We have already confirmed that an absorption edge of YSZ substrate locates at $\sim$4.8 eV~\cite{Ito2021} which is far away from the displayed range. 
The onset of sharp increase in $\alpha$ is ranging from  $\sim$2.6 to $\sim$3.4 eV.
The lowest absorption-edge energy is observed in {\SNO}, followed in order by {\STO}, {\PNO}, and {\PTO}.
As we have already reported in the previous work~\cite{Ito2021}, these absorption edges are plausibly assignable to a transition from the hybridized \textit{A}-n\textit{s} and O-2\textit{p} valence band to \textit{B}-n\textit{d} conduction band.
It is worth mentioning that precise optical measurements for thin-film samples enable us to capture the aforementioned trends, while systematic variation of band gap has been reported through reflectance measurements for bulk Sn$_2$(Nb$_{1-x}$Ta$_x$)$_2$O$_7$ samples~\cite{Kikuchi2017}. 
These results of our films confirm that the hybridized band is the common feature in this class of compounds as predicted by the first-principles band calculations~\cite{Hase2018_1,Hase2018_2,Hase2019,Hase2020}.

\begin{figure*}[h]
	\includegraphics[width=12.4cm]{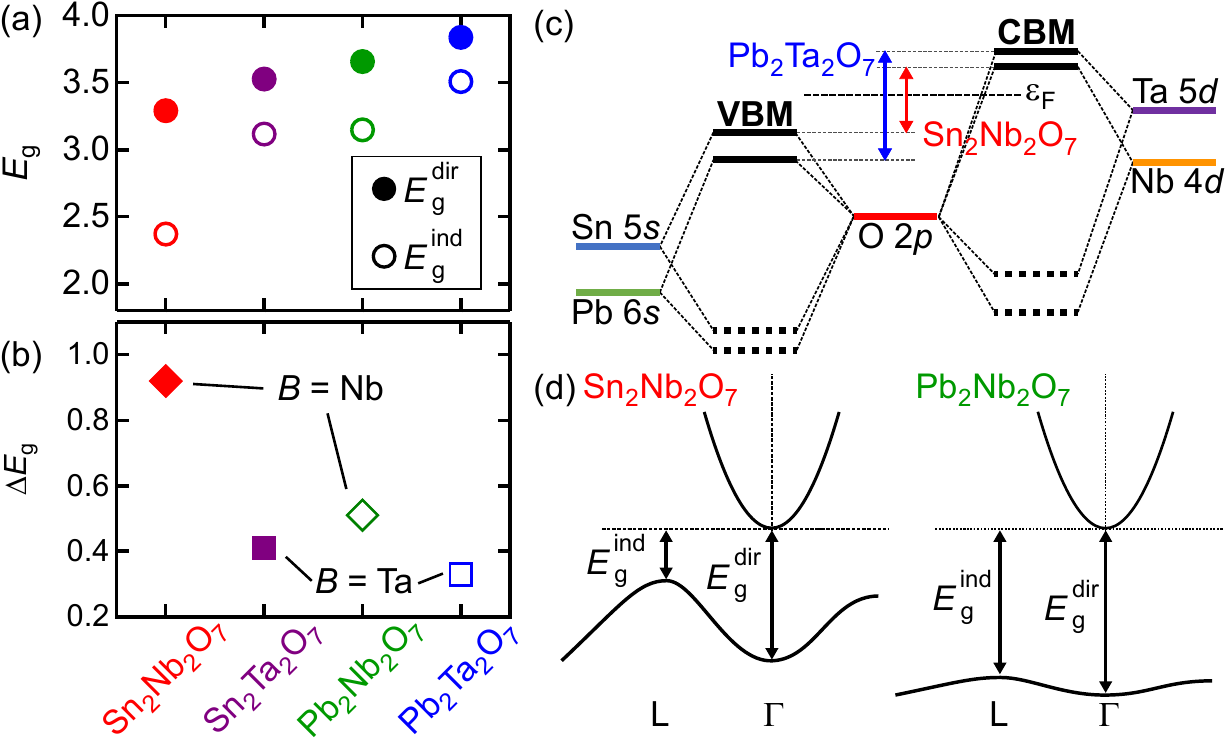}
	\colorcaption{\label{Eg} 
		(a)  Direct and indirect band gaps ($E^{\rm{dir}}_{\rm{g}}$ and $E^{\rm{ind}}_{\rm{g}}$) of {\SNO},  {\STO}, {\PNO}, and {\PTO}  films deduced from the Tauc plots in Figs.~\ref{opt}(b) and~\ref{opt}(c).
		(b) Difference between direct and indirect band gaps ($\Delta E_{\rm{g}}\equiv E^{\rm{dir}}_{\rm{g}} - E^{\rm{ind}}_{\rm{g}}$) of the films.
		(c) Schematic energy level diagram based on molecular orbital method considering the energy levels of Sn-5\textit{s}, Pb-6\textit{s}, Nb-4\textit{d}, Ta-5\textit{d}, and O-2\textit{p} orbitals.
		Dashed and solid lines indicate the bonding and anti-bonding orbitals, respectively. 
		Valence band maximum (VBM) and conduction band minimum (CBM) consist of anti-bonding orbitals between \textit{A}-n\textit{s} and O-2\textit{p}, and anti-bonding orbitals between \textit{B}-n\textit{d} and O-2\textit{p} orbitals, respectively.
		(d) Schematic band structures of {\SNO} and {\PNO}. 
			}
\end{figure*}

We then focus on the optical band gap ($E_{\rm{g}}$)  determined from the Tauc plots with considering direct and indirect  allowed transitions.
In Figs.~\ref{opt}(b) and \ref{opt}(c), $(\alpha h\nu)^{2}$ and $(\alpha h\nu)^{1/2}$ are plotted as a function of $h\nu$, where direct and indirect band gaps ($E^{\rm{dir}}_{\rm{g}}$ and $E^{\rm{ind}}_{\rm{g}}$) can be deduced from the intercept of the dashed line with the horizontal axis.  
Figure~\ref{Eg}(a) summarizes the deduced $E^{\rm{dir}}_{\rm{g}}$ and $E^{\rm{ind}}_{\rm{g}}$ of all the films. 
As in the case of absorption edge, the narrowest band gap, both direct and indirect, is observed in {\SNO}, followed in order by {\STO}, {\PNO}, and {\PTO}.
In accordance with the observed order of the band gaps, a schematic energy level diagram of each cation around Fermi level ($\epsilon_F$) is illustrated in the inset of Fig.~\ref{opt}(b). 

This tendency can be further understood by considering the energy levels of atomic orbitals for Sn-5\textit{s}, Pb-6\textit{s}, Nb-4\textit{d}, Ta-5\textit{d}, and O-2\textit{p} as shown in Fig.~\ref{Eg}(c). 
This diagram is based on the following consideration:
(i) in this class of compounds, band gap is the energy difference between two anti-bonding orbitals; one is \textit{A}-n$s$ and O-2\textit{p}, and the other is \textit{B}-n$d$ and O-2\textit{p}, the former of which forms VBM and the latter does CBM, respectively,
(ii) energy levels of the orbitals of the constituent elements are in the order of Pb-6\textit{s}, Sn-5\textit{s}, O-2\textit{p}, Nb-4\textit{d}, and Ta-5\textit{d} from deeper to shallower energy.
The order that Pb-6\textit{s} is deeper than Sn-5\textit{s} might seem unnatural because elements with the larger principal quantum number usually have  shallower energy levels if their orbital angular momentum is the same, as in the case of the relationship between  Nb-4\textit{d}, and Ta-5\textit{d} orbitals.
However, this seemingly unnatural ``inversion'' has been actually observed in Pb$_x$Sn$_{1-x}$O by the fact that its band gap becomes larger by increasing the Pb doping ratio \textit{x}, which is also supported by a DFT calculation~\cite{Liao2015}.  
This phenomenon is explained by both orbital penetration close to nucleus and relativistic effects originating from Pb-6\textit{s} atomic orbital~\cite{Liao2015}, and thus this can be the case for pyrochlore oxides in this study as well.

Another intriguing observation is the difference between direct and indirect band gaps ($\Delta E_{\rm{g}}\equiv E^{\rm{dir}}_{\rm{g}} - E^{\rm{ind}}_{\rm{g}}$)  summarized in Fig.~\ref{Eg}(b). 
Here, we can see two general trends: (i) when \textit{B}-site is common, $\Delta E_{\rm{g}}$ becomes smaller in \textit{A} = Pb compounds than in \textit{A} = Sn ones, (ii) when \textit{A}-site is common, $\Delta E_{\rm{g}}$ becomes smaller in  \textit{B} = Ta compounds than in \textit{B} = Nb ones.
According to the previously reported theoretical calculations for this class of compounds~\cite{Hase2018_1,Hase2019,Hase2020}, it is revealed that VBM (CBM) becomes flattened when Sn- (Nb-)site is substituted by Pb (Ta).  
Especially in $A_2$Nb$_2$O$_7$, it is clear that the direct and indirect transitions correspond to those from $\Gamma$ to $\Gamma$  points and from $L$ to $\Gamma$  points, respectively~\cite{Hase2018_1,Hase2019}, as schematically illustrated in Fig.~\ref{Eg}(d). 
Due to this, $\Delta E_{\rm{g}}$ can be a good indicator of the ``flatness'' in valence band; smaller $\Delta E_{\rm{g}}$ indicates a flatter band with a smaller dispersion in VBM.
On the other hand, for $A_2$Ta$_2$O$_7$, CBM is also expected to be flattened and the energy levels of $\Gamma$ and $L$ points in CBM become closer~\cite{Hase2020}, which makes the simple interpretation of $\Delta E_{\rm{g}}$ in $A_2$Nb$_2$O$_7$ no longer valid.  
Because of this more complicated CBM structure, it is not straightforward to relate the reduction in $\Delta E_{\rm{g}}$ with the reduction of the band dispersion in VBM, which might be also the reason why  $\Delta E_{\rm{g}}$ is larger in {\PNO} than in {\STO}.
Nonetheless, considering that VBM should reflect the nature of the \textit{A}-site element as mentioned above, we speculate that  {\PTO} has a less dispersive VBM than {\STO}. 
Together with the energy  level diagram in Fig.~\ref{Eg}(c), we conclude that Pb substitution in Sn$_2$$B_2$O$_7$ has the following two effects on the band structure: (i) enlarges the band gap, (ii) reduces the band dispersion in VBM.

In summary, epitaxial {\FBO} (111) films are grown on YSZ (111) substrates by pulsed laser deposition.
Because PbO has higher vapor pressure than SnO, higher O$_2$ pressure is required to fabricate Pb$_2$$B_2$O$_7$ compared with the case of Sn$_2$$B_2$O$_7$.
Optical measurements reveal that the absorption edge appears at between $\sim$2.6 and $\sim$3.4 eV in our films, which is suggestive of the existence of the hybridized \textit{A}-n\textit{s} and O-2\textit{p} band as predicted by theoretical calculations.  
Both direct and indirect band gaps of {\SNO} are the smallest,  followed in order by those of {\STO}, {\PNO}, and {\PTO}.
This tendency can be interpreted by taking into account the energy levels of Sn-5\textit{s}, Pb-6\textit{s}, Nb-4\textit{d}, and Ta-5\textit{d} atomic orbitals.
By comparing the difference between direct and indirect band gaps, it is clarified that  Pb$_2$$B_2$O$_7$ tends to have smaller valence band dispersion compared with  Sn$_2$$B_2$O$_7$.
These general trends will be helpful for band engineering to materialize numbers of emergent phenomena expected in this class of compounds.
In this regard, precise stoichiometry control should be required, which remains as future work.

This work was supported by the Japan Science and Technology Agency Core Research for Evolutional Science
and Technology (JST CREST) (Grant No. JPMJCR16F1), by JSPS Grant-in-Aid for Early-Career Scientists No. JP20K15168, and by The Murata Science Foundation.

The data that support the findings of this study are available from the corresponding author upon reasonable request.
\bibliography{FBO}
\textbf{}
\newline
\newline
\textbf{FIGURES}

\newpage
\textbf{}
\newline\newline
\newline
\newline
\newline
\newline

\newpage
\textbf{}
\newline\newline
\newline
\newline
\newline
\newline

\end{document}


\begin{center}
\textbf{\large Supplemental Materials:\\Trends in Bandgap of Epitaxial \textit{A}$_2$\textit{B}$_2$O$_7$ (\textit{A} = Sn, Pb; \textit{B} = Nb, Ta) Films Fabricated by Pulsed Laser Deposition}
\end{center}
\renewcommand{\thepage}{\arabic{page}/{4}}
\setcounter{equation}{0}
\setcounter{figure}{0}
\setcounter{table}{0}
\setcounter{page}{1}
\makeatletter
\renewcommand{\theequation}{S\arabic{equation}}
\renewcommand{\thefigure}{S\arabic{figure}}
\renewcommand{\thetable}{S\arabic{table}}
\renewcommand{\bibnumfmt}[1]{[S#1] }
\renewcommand{\citenumfont}[1]{S#1}

\section{Reciprocal space mapping and azimuthal scans}
\begin{figure*}[h]
	\includegraphics[width=15cm]{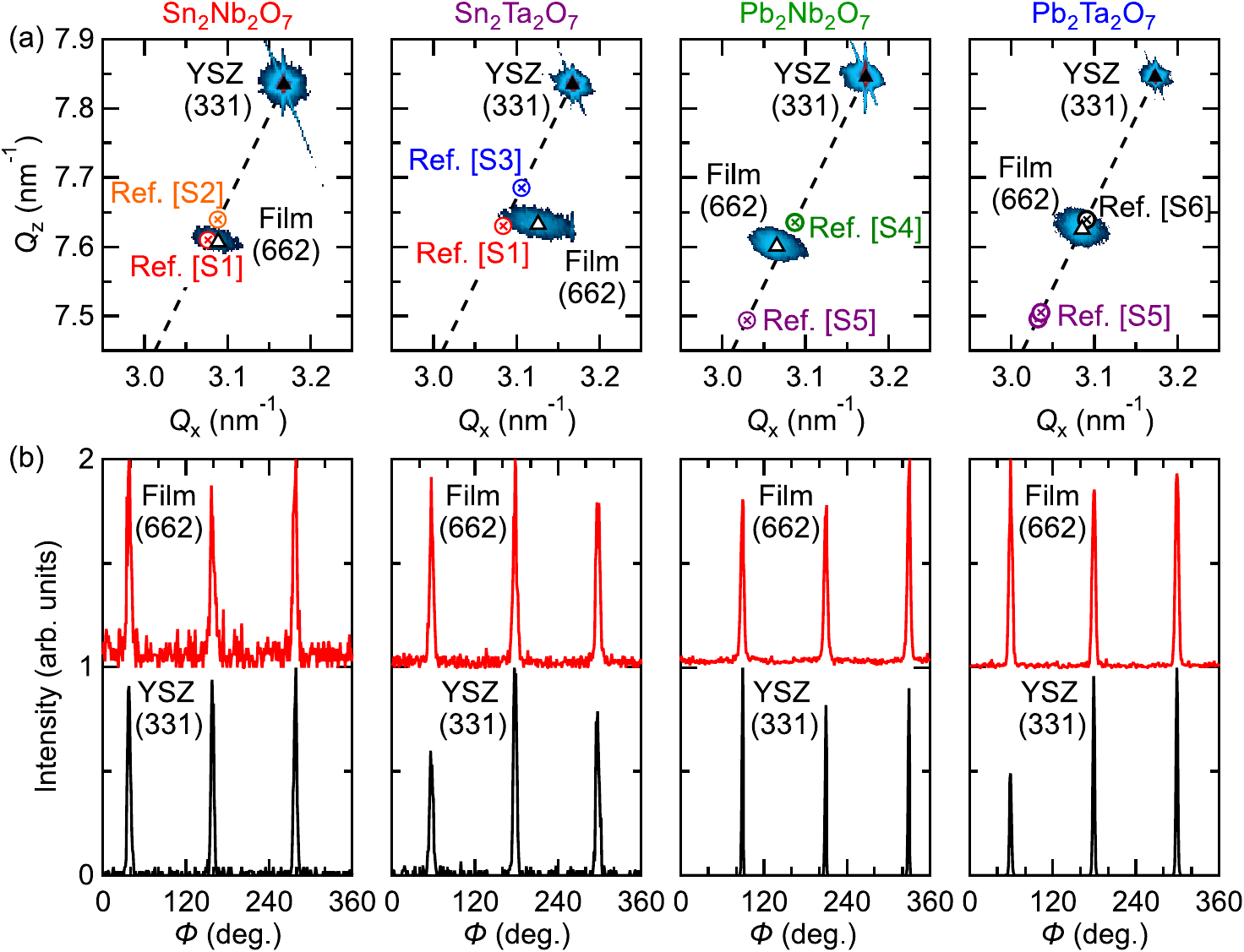}
	\colorcaption{\label{RSM}
		(a) Reciprocal space mapping (RSM) around pyrochlore (662) and YSZ (331) peaks for {\SNO},  {\STO}, {\PNO}, and {\PTO} films. 
		Peaks of the films and YSZ substrate are indicated by black open  and closed triangles, respectively.
		The data for polycrystalline bulk samples are indicated by open circles with cross as summarized in Table~\ref{tab:lattice}~\cite{Kikuchi2017,Ubic2002,Nalini2001,Scott1982,Cruz2001,Brisse1972}.  
		Dotted line indicates the relaxation line where fully relaxed cubic film peak is supposed to be located. 
		(b) Azimuthal ($\phi$) scan around pyrochlore (662) (top panels, red) and YSZ (331) (bottom panels, black) peaks.
	}
\end{figure*}

Figure~\ref{RSM}(a) shows reciprocal space mappings (RSM) around YSZ (331) and pyrochlore (662) peaks for {\SNO},  {\STO}, {\PNO}, and {\PTO} films, together with the data for previously reported bulk samples~\cite{Kikuchi2017,Cruz2001,Brisse1972,Ubic2002,Nalini2001,Scott1982}.
Details of the bulk samples are summarized in  Table~\ref{tab:lattice}, where the lattice mismatch with YSZ substrate is calculated in comparison with the twice of lattice constant of YSZ (5.139 \AA). 
We have to note that even in the bulk samples, the reported lattice constants have considerable variations because of crystalline defects such as off-stoichiometry and anti-site defects (\textit{A}-site element locates at \textit{B}-site as $A^{4+}$ instead of locating at \textit{A}-site as $A^{2+}$)~\cite{Kikuchi2017,Cruz2001,Nalini2001}.
These defects might be present in our films as well, but it is challenging to detect because of the difficulty in chemical composition analysis as we mention in the main text.
The dotted lines in RSM indicate the relaxation line where fully relaxed cubic film peak is supposed to be located.
Although the peak positions of Pb$_2$$B_2$O$_7$ films are on the relaxation line, those of Sn$_2$$B_2$O$_7$ films are slightly shifted toward the substrate peak, indicating that Sn$_2$$B_2$O$_7$ films remain weakly locked to the substrate maybe because of relatively smaller lattice mismatch compared with Pb$_2$$B_2$O$_7$.  
Azimuthal ($\phi$) scans for these peaks are shown in Fig.~\ref{RSM}(b), exhibiting a three-fold rotational symmetry as expected from the crystal structure.  

\newpage
\textbf{}
\newline
\newline\newline
\newline\newline
\newline

\begin{table}[h]
\caption{\label{tab:lattice}Lattice constants of bulk samples}
\begin{ruledtabular}
\begin{tabular}{cccc}
\textbf{Compound} & \textbf{Lattice constant (\AA)} &

\begin{tabular}{c}
 \textbf{Mismatch}\\\textbf{with YSZ (\%)}
\end{tabular}
& \textbf{Ref.} \\ \hline

Sn$_{1.48}$(Nb$_{1.74}$Sn$_{0.26})$O$_{6.35}$& 10.580 & 2.94 & \cite{Kikuchi2017}\\
Sn$_{1.34}$(Nb$_{1.68}$Sn$_{0.32})$O$_{6.18}$& 10.539 & 2.54 & \cite{Cruz2001}\\

Sn$_{1.44}$(Ta$_{1.77}$Sn$_{0.23})$O$_{6.43}$ & 10.555 & 2.70 & \cite{Kikuchi2017}\\
Sn$_{2}$Ta$_{2}$O$_{7}$ & 10.475 & 1.97 & \cite{Brisse1972}\\

Pb$_{1.5}$Nb$_{2}$O$_{6.5}$ & 10.562 & 2.75 & \cite{Ubic2002}\\
Pb$_{2}$(Nb$_{1.51}$Sn$_{0.49})$O$_{6.30}$ & 10.762 &  4.71 & \cite{Nalini2001}\\

Pb$_{1.5}$Ta$_{2}$O$_{6.5}$ & 10.555 & 2.7 & \cite{Scott1982}\\
Pb$_{2}$(Ta$_{1.4}$Pb$_{0.6})$O$_{6.21}$ & 10.744 & 4.53 & \cite{Nalini2001}\\
Pb$_{2}$(Ta$_{1.25}$Pb$_{0.75})$O$_{6.57}$ & 10.757 & 4.66 & \cite{Nalini2001}\\

\end{tabular}
\end{ruledtabular}

\end{table}

\newpage
\textbf{}
\newline
\newline
\newline

\clearpage

\section{Atomic force microscope images}
\begin{figure*}[h]
	\includegraphics[width=11.2cm]{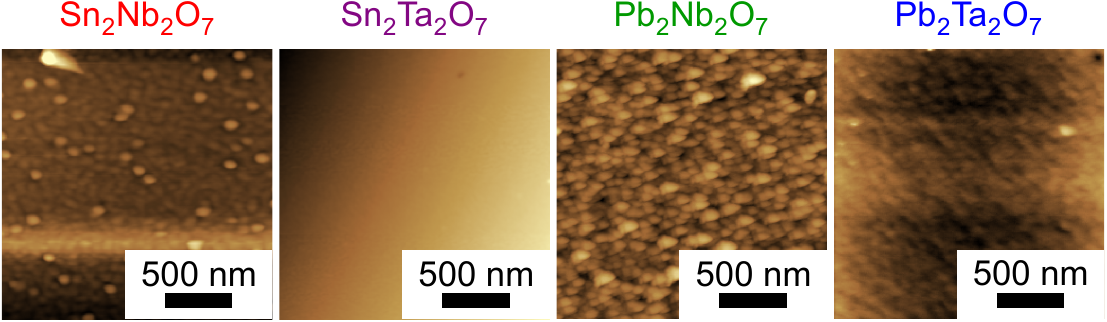}
	\colorcaption{\label{AFM}
	AFM images of the films.
	}
\end{figure*}

Figure~\ref{AFM} summarizes atomic force microscope  (AFM) images of {\SNO},  {\STO}, {\PNO}, and {\PTO} films, the route-mean-square (RMS) roughness of which are 1.0, 0.3, 1.2, and 0.4 nm, respectively. 
\\

\bibliography{FBO}